\documentclass[12pt]{article}

\usepackage{sbc-template}
\usepackage{graphicx,url}
\usepackage{tikz}
\usepackage{helvet}
\usepackage[utf8]{inputenc}
\usepackage[T1]{fontenc}
\usepackage{booktabs}
\usepackage{xspace}
\usepackage{microtype}
\usepackage[cc]{titlepic}
\usepackage[hidelinks]{hyperref}
\usepackage{fancyhdr}

\fancypagestyle{firstpage}{%
    \fancyhf{}%
    \fancyhead[L]{%
        \includegraphics[height=1.2cm]{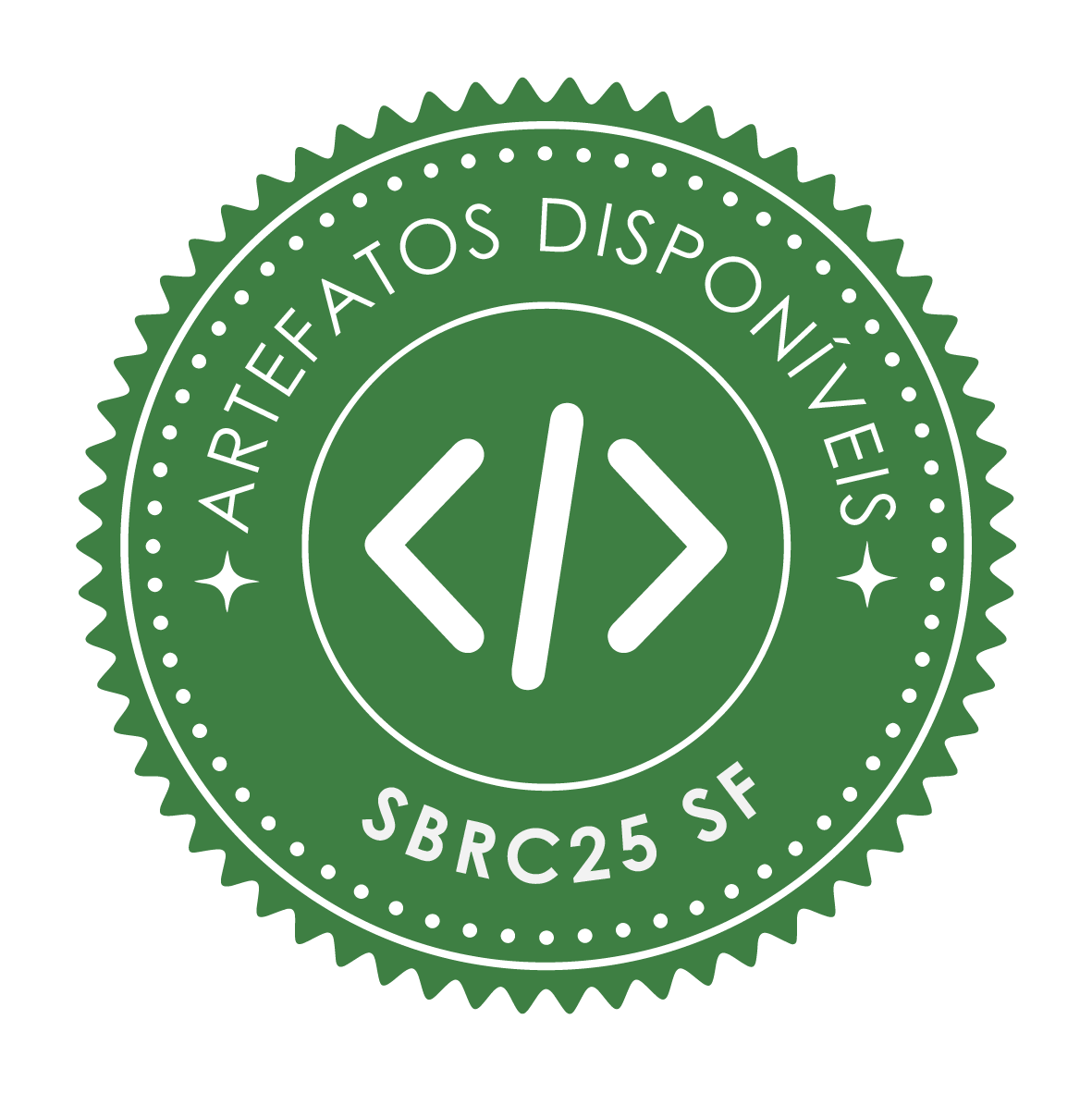}%
        \hspace{0.15cm}%
        \includegraphics[height=1.2cm]{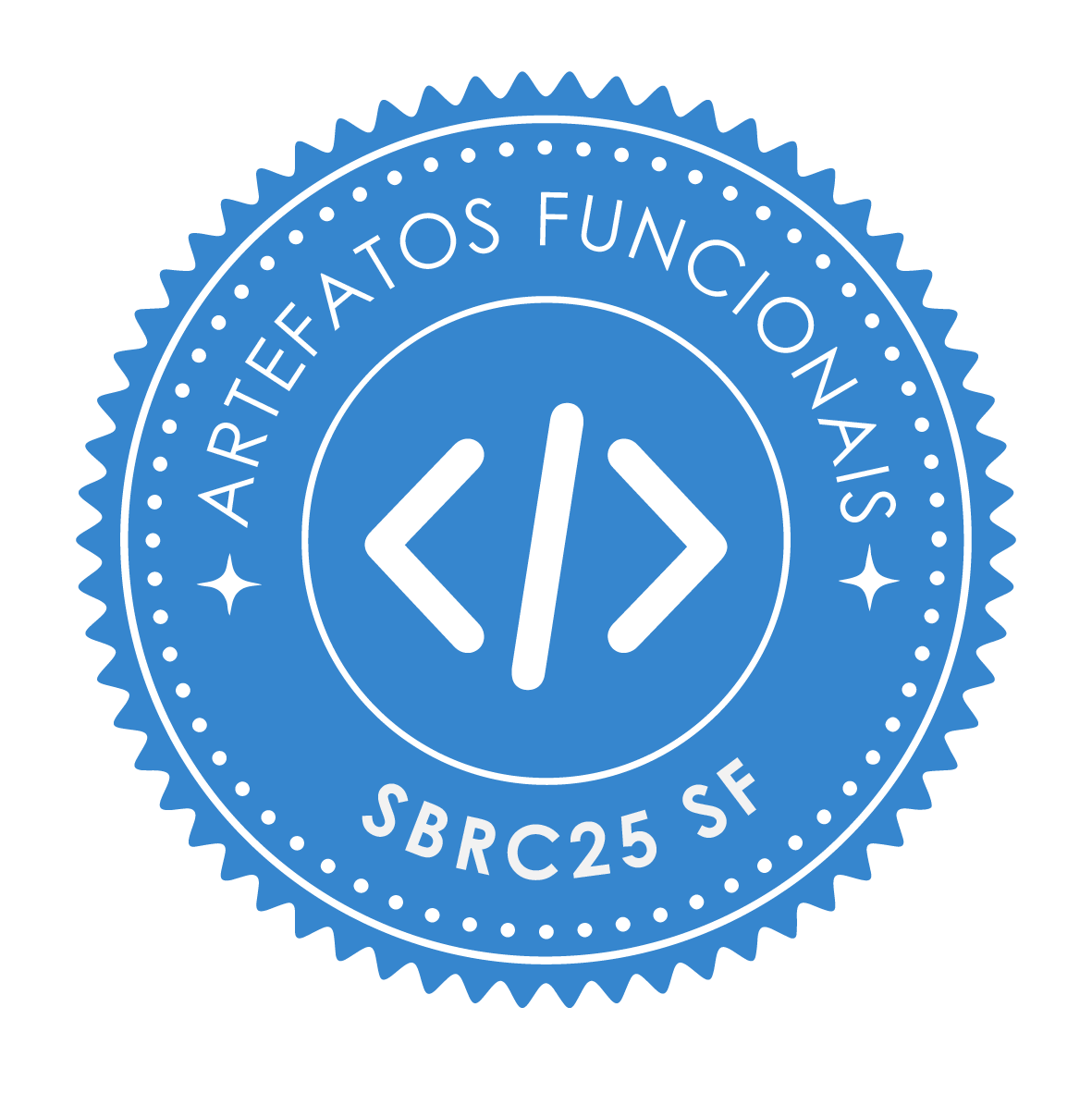}%
        \hspace{0.15cm}%
        \includegraphics[height=1.2cm]{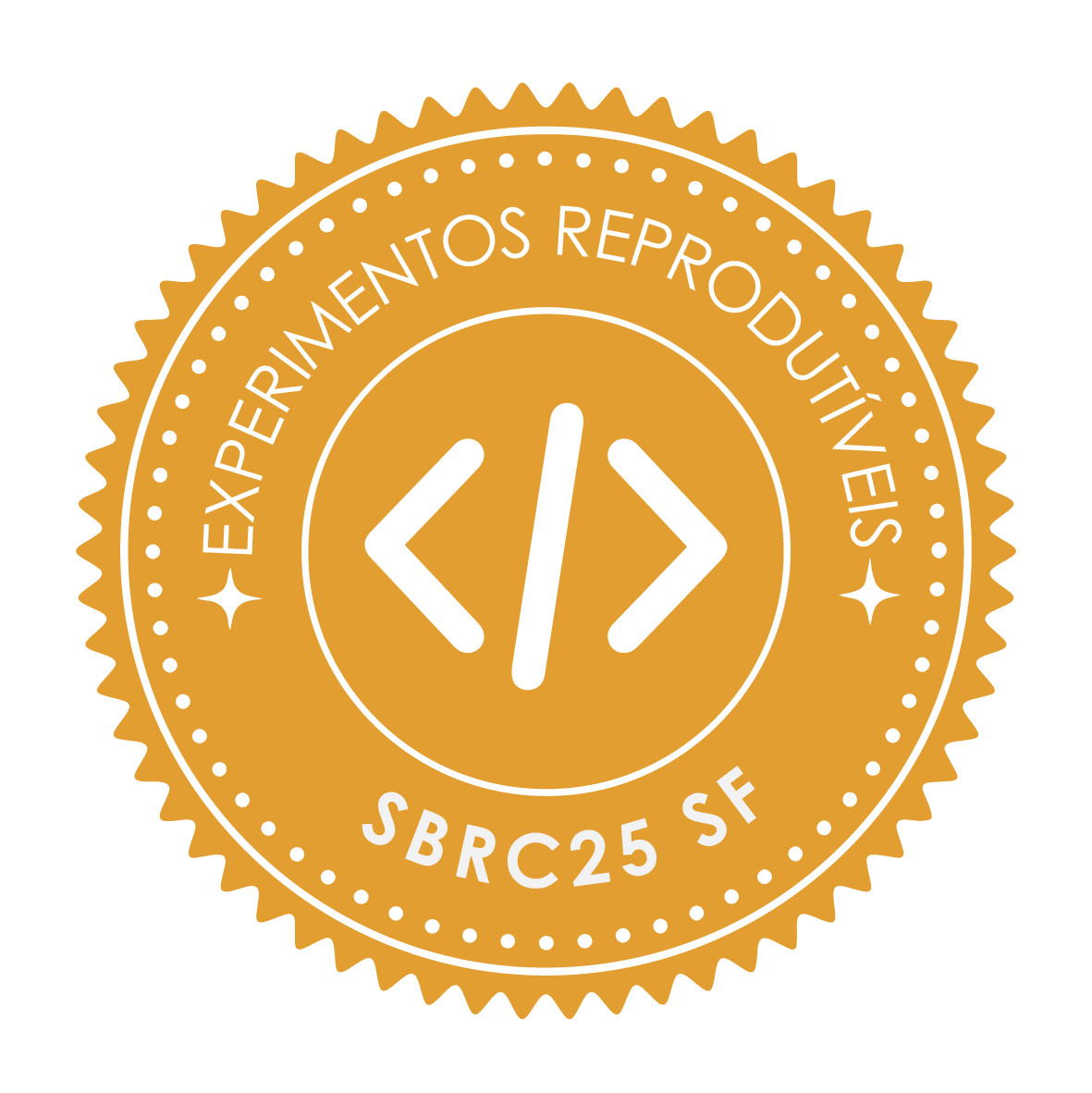}%
    }%

    \fancyhead[R]{%
        \href{https://doi.org/10.5753/sbseg.2026.29056}{%
            \includegraphics[width=2cm]{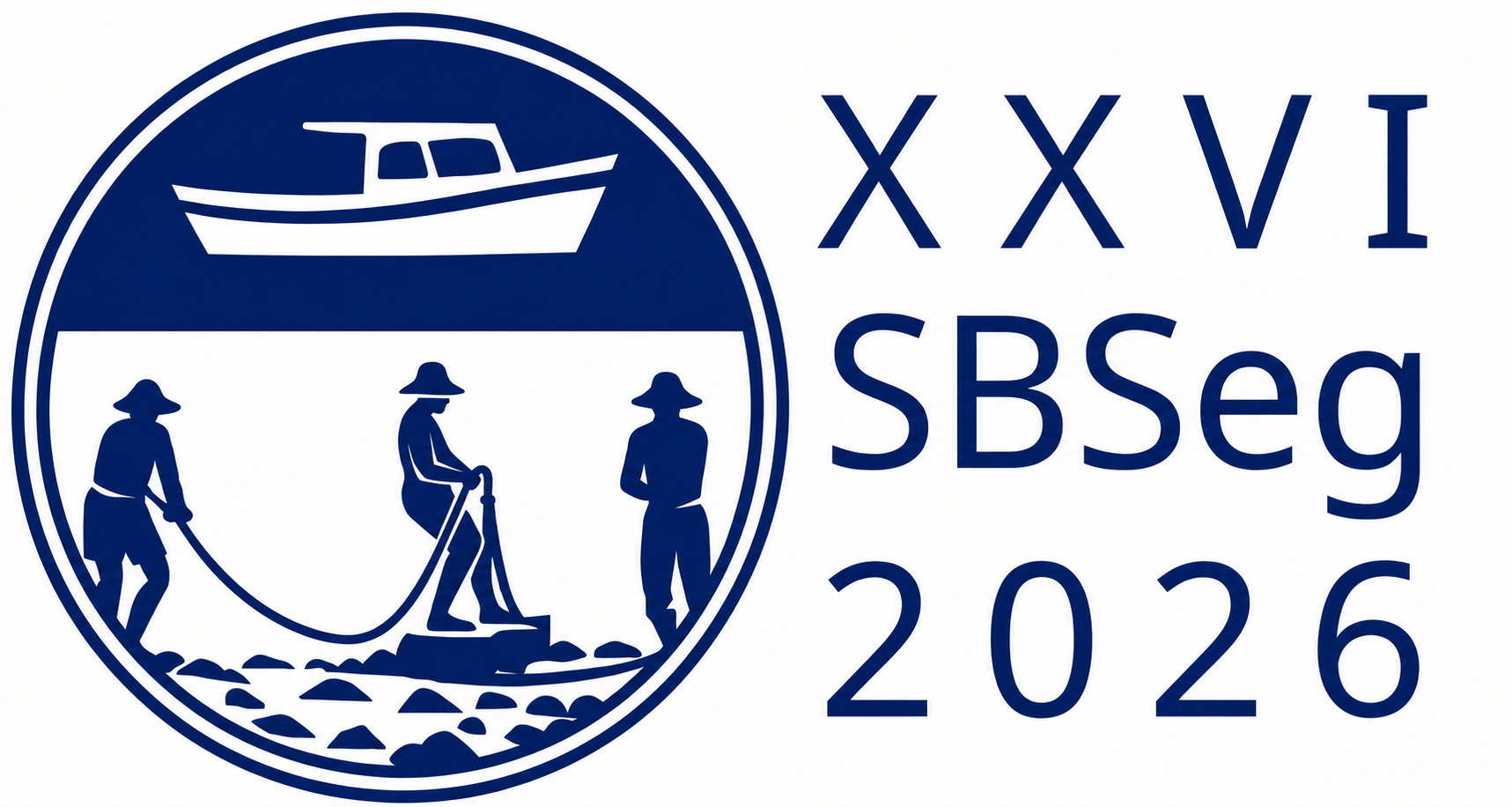}%
        }%
    }%
}

\usetikzlibrary{arrows.meta,positioning,calc,backgrounds}
\definecolor{tvink}{HTML}{1F2933}
\definecolor{tvmuted}{HTML}{7B8794}
\definecolor{tvline}{HTML}{C7D0DA}
\definecolor{tvaccent}{HTML}{24527A}
\definecolor{tvaccentbg}{HTML}{F3F7FA}
\definecolor{tvsoft}{HTML}{FAFBFC}
\definecolor{tvfail}{HTML}{9A3333}
\definecolor{tvfailbg}{HTML}{FFF9F9}
\RequirePackage{silence}
\newcommand{\tv}{\textit{TopVenues}\xspace}

\title{TopVenues: A Reproducible Corpus and Tooling Substrate for Cybersecurity Literature Reviews}

\author{Sidnei Barbieri\inst{1}, \'{A}gney Lopes Roth Ferraz\inst{1} and Louren\c{c}o Alves Pereira J\'unior\inst{1}}

\address{Divisão de Ciência da Computação -- Instituto Tecnológico de Aeronáutica
  (ITA)\\
  Praça Marechal Eduardo Gomes, 50 -- 12228-900 -- São José dos Campos -- SP -- Brasil
\email{sidneisb@ita.br, roth@ita.br and ljr@ita.br}
}

\begin{document}

\maketitle
\thispagestyle{firstpage}

\begin{abstract}
Cybersecurity literature reviews require a reproducible denominator: the set of papers that a protocol includes before screening and synthesis begin. Today, that denominator is often reconstructed from publisher portals, bibliographic indices, and scholarly application programming interfaces (APIs) whose coverage, formats, and query semantics change over time. This paper presents \tv, an open-source system that materializes corpus construction as a versioned research artifact. \tv declares a venue and year scope, uses the DBLP Computer Science Bibliography (DBLP) as its primary metadata source, enriches records with abstracts and BibTeX entries via open scholarly APIs and publisher-specific extractors, and stores the results in a monotonic SQLite snapshot, accessible via a command-line interface (CLI), a web interface, and export paths for review workflows. The May 2026 snapshot contains 9,925 bibliographic records from 11 configured security-focused, security-relevant adjacent, and survey venues, with 99.86\% abstract coverage and 99.99\% BibTeX coverage; keyword search over the full corpus completes in under 31\, ms, and a 252-test suite validates the data-integrity invariants. The fixed denominator also enables repeatable measurement: 29.2\% of 2024 to 2025 papers from the four top-ranked security conferences in our scope appear as arXiv preprints, with a median of five months before publication, and a prior-author-track-record filter yields a 16.5$\times$ relative risk (2.5$\times$ conventional lift) at 90\% recall for triaging preprints that later appear in the same venue set. \tv links corpus construction to auditable cybersecurity measurement by making the corpus itself executable, inspectable, and citable. The artifact is available at \url{https://github.com/sidneibarbieri/topVenues}.
\end{abstract}

\section{Introduction}
\label{sec:intro}

Cybersecurity is one of the most prolific fields in computer science. The four security conferences classified in the A* tier, the highest tier, by the Computing Research and Education (CORE) ranking (USENIX Security, ACM CCS, IEEE S\&P, and NDSS), account for nearly 6,000 papers in our corpus since 2019. Adding adjacent conferences and survey journals pushes the current review surface above 9,900 records. For a researcher writing a survey, positioning a new system, or tracking an emerging topic, such as Large Language Model (LLM) security, the first scientific question is often blocked by a more basic one: \emph{What is the population of papers being reviewed?}

The usual answer is procedural, not reproducible. A researcher queries DBLP, ACM Digital Library, IEEE Xplore, USENIX, NDSS, publisher portals, and scholarly application programming interfaces (APIs), then reconciles results in a spreadsheet or reference manager. This workflow can be documented, but it does not produce a stable denominator. Portal ranking changes, late DBLP updates, missing abstracts, rate limits, and inconsistent BibTeX exports all affect what is recovered. The resulting corpus is often treated as a temporary byproduct of the review rather than as a scientific object that another researcher can inspect, rerun, and extend.

This denominator gap matters because literature reviews now serve a purpose beyond background sections. They drive research ideation, artifact positioning, technology monitoring, and meta-scientific measurements across the security community. Systematic literature review (SLR) methodology separates protocol design, search, selection, quality assessment, extraction, and synthesis~\cite{kitchenham2007slr,page2021prisma,rethlefsen2021prismas}; screening tools such as Rayyan and ASReview improve collaborative triage and prioritization~\cite{ouzzani2016rayyan,van2021asreview}. These methods improve review discipline after a search process is specified, but they do not make the candidate set itself reproducible.

Cybersecurity reviews highlight the importance of this missing layer. Surveys on intrusion detection and targeted attacks must reconcile heterogeneous venues, terminology, and evidence sources~\cite{khraisat2019,luh2017}. Cyber Threat Intelligence (CTI) work spans academic papers, vendor reports, public feeds, and sharing communities~\cite{shu2018threatintelligence,li2019tealeaves,bouwman2020commercialti,bouwman2022helpinghands,barbieri2025searching}. In these settings, the corpus is not a clerical detail; it is the boundary of the claim. A survey that silently changes its denominator changes the evidence available to the reader. A reproducible cybersecurity review, therefore, needs the same kind of explicit protocol state for corpus construction that SLR guidelines already require for search documentation, study selection, and synthesis.

\tv addresses this missing layer by making the corpus a versioned artifact. It defines a configurable publication scope, builds a DBLP-backed record set, enriches it with abstracts and BibTeX entries from multiple sources, and preserves the results as a monotonic SQLite snapshot. The snapshot is immediately usable through a command-line interface (CLI) and web interface, and its scientific property is the denominator: every count, query, export, and measurement is tied to a declared denominator that can be reproduced from the released artifact.

This paper makes five contributions. It formulates corpus construction as a distinct reproducibility problem, presents the \tv architecture, describes the reproducibility controls that make the artifact practical to execute, releases and characterizes a 9,925-record venue-bounded corpus for cybersecurity literature reviews, and uses that corpus as a denominator for two measurements: a 29.2\% arXiv early-signal rate and a 16.5$\times$ relative risk (2.5$\times$ conventional lift) from a scientific-readiness filter. The work falls under data management and cybersecurity systems, within the broader theme of security-oriented data management and measurement.

\section{Problem and Requirements}
\label{sec:problem}

We distinguish \emph{corpus construction} from \emph{corpus use}. Corpus construction assembles the population of papers that satisfy a declared scope: venues, years, paper classes, metadata completeness, and refresh policy. Corpus use begins only after that population exists: screening, annotation, synthesis, querying, exporting, and measurement. Most review tools operate in the second phase. \tv targets the first, where an implicit search procedure must be made explicit and reusable.

This framing yields six requirements. The scope must be curated and machine-readable (R1) because venue selection determines the literature population and should be configured via versioned configuration rather than an unrecorded portal query. The metadata path must rely on open sources whenever possible (R2), so reviewers can reproduce the collection logic without institutional credentials. The corpus must be abstract-complete enough for topic filtering and semantic triage (R3), since title-only search misses much of the evidence that review work depends on. Refreshes must be monotonic (R4): once an abstract or BibTeX entry has been recovered, a later null response from another source must not erase it. The corpus must export directly to the tools researchers already use (R5), including BibTeX, comma-separated values (CSV), and JavaScript Object Notation (JSON). Finally, the artifact must be installable, documented, and testable from a fresh clone (R6), so that the main claims can be validated without publisher subscriptions or live scholarly APIs.

These requirements define a boundary condition for a research artifact. The system does not need to automate scholarly judgment; it must preserve the inputs to that judgment. A reviewer should be able to ask why a record is present, which stable identifiers and source links support it, whether the same snapshot can be reconstructed, and which records remain outside abstract-aware analysis. Those questions are the difference between a search result and a reproducible denominator.

\section{Architecture}
\label{sec:architecture}

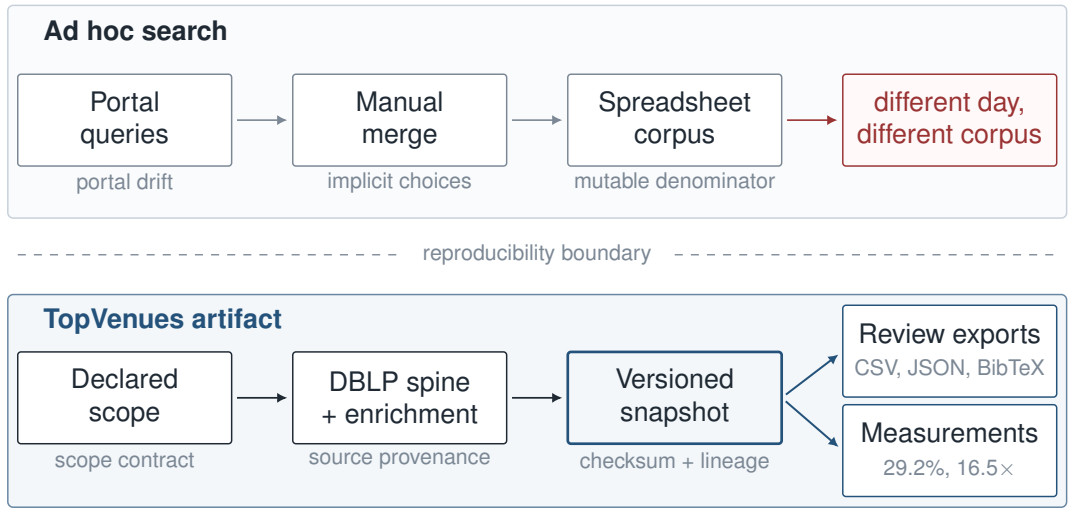
\begin{figure}[ht]
\centering
\resizebox{\columnwidth}{!}{\begin{tikzpicture}[
  font=\sffamily\footnotesize,
  >={Latex[length=3.8pt,width=3.8pt]},
  lane/.style={draw=tvline,line width=.55pt,fill=tvsoft,rounded corners=2pt},
  rowtitle/.style={font=\sffamily\footnotesize\bfseries,text=tvink,anchor=west,align=left},
  box/.style={draw=tvink,line width=.55pt,fill=white,rounded corners=1.2pt,
              minimum width=28mm,minimum height=12mm,align=center,inner xsep=1.2pt,inner ysep=0pt,
              outer sep=0pt,text=tvink},
  dim/.style={box,draw=tvmuted,text=tvink},
  key/.style={box,draw=tvaccent,line width=1pt,fill=tvaccentbg},
  fail/.style={box,draw=tvfail,fill=tvfailbg,text=tvfail},
  resultbox/.style={box,draw=tvaccent,text=tvink},
  note/.style={font=\sffamily\scriptsize,text=tvmuted,align=center,inner sep=0pt},
  flow/.style={->,line width=.65pt,draw=tvink,shorten >=2pt,shorten <=2pt},
  dflow/.style={->,line width=.6pt,draw=tvmuted,shorten >=2pt,shorten <=2pt},
  aflow/.style={->,line width=.75pt,draw=tvaccent,shorten >=2pt,shorten <=2pt},
]
\path[use as bounding box] (0.02,-3.90) rectangle (13.88,2.85);
\draw[lane] (0.02,0.00) rectangle (13.88,2.78);
\draw[lane,draw=tvaccent!70,fill=tvaccentbg] (0.02,-3.78) rectangle (13.88,-1.00);
\node[rowtitle] at (0.34,2.45) {Ad hoc search};
\node[rowtitle,text=tvaccent] at (0.34,-1.35) {TopVenues artifact};
\node[dim] (portal) at (1.55,1.28) {Portal\\queries};
\node[dim] (merge)  at (5.15,1.28) {Manual\\merge};
\node[dim] (sheet)  at (8.75,1.28) {Spreadsheet\\corpus};
\node[fail] (drift) at (12.35,1.28) {different day,\\different corpus};
\draw[dflow] (portal) -- (merge);
\draw[dflow] (merge) -- (sheet);
\draw[->,line width=.6pt,draw=tvfail,shorten >=2pt,shorten <=2pt] (sheet) -- (drift);
\node[note,below=.85mm of portal] {portal drift};
\node[note,below=.85mm of merge] {implicit choices};
\node[note,below=.85mm of sheet] {mutable denominator};
\draw[dashed,line width=.5pt,draw=tvmuted] (0.15,-0.48) -- (5.15,-0.48);
\draw[dashed,line width=.5pt,draw=tvmuted] (8.75,-0.48) -- (13.75,-0.48);
\node[fill=white,inner xsep=5pt,font=\sffamily\scriptsize,text=tvmuted] at (6.95,-0.48)
  {reproducibility boundary};
\node[box] (scope) at (1.55,-2.35) {Declared\\scope};
\node[box] (spine) at (5.15,-2.35) {DBLP metadata\\+ enrichment};
\node[key] (snap)  at (8.75,-2.35) {Versioned\\snapshot};
\node[resultbox] (review) at (12.35,-1.74) {Review exports\\{\scriptsize\color{tvmuted}CSV, JSON, BibTeX}};
\node[resultbox] (measure) at (12.35,-3.04) {Measurements\\{\scriptsize\color{tvmuted}29.2\%, RR 16.5$\times$}};
\draw[flow] (scope) -- (spine);
\draw[flow] (spine) -- (snap);
\draw[aflow] (snap.east) -- (review.west);
\draw[aflow] (snap.east) -- (measure.west);
\node[note,below=1.1mm of scope] {scope contract};
\node[note,below=1.1mm of spine] {source evidence};
\node[note,below=1.1mm of snap] {checksum + lineage};
\end{tikzpicture}}
\caption{Ad hoc search rebuilds a mutable corpus, while \tv produces a versioned denominator for exports and measurements.}
\label{fig:arch}
\end{figure}

\tv is organized around a simple principle: the corpus is the persistent scientific object, and every tool is either a constructor, validator, or consumer of that object. Figure~\ref{fig:arch} contrasts this model with the common manual workflow. In the ad hoc model, portal queries reconstruct a corpus each time, and the denominator remains entangled with search semantics, deduplication choices, and spreadsheet state. In \tv, the declared scope enters a controlled construction boundary: DBLP supplies the primary metadata, enrichment sources supply abstracts and BibTeX, and the monotonic SQLite snapshot becomes the object that review workflows and measurement studies consume.

The construction pipeline has four stages, each of which can be executed independently for auditing or chained into a full refresh. The download stage fetches DBLP JSON records for each configured source and year range, caching them in a per-source, per-year layout so that incremental runs retrieve only missing proceedings. Consolidation then normalizes source names, collapses repeated ingestion of the same DBLP record by its API identifier, and writes records to SQLite. Distinct DBLP identities are deliberately retained, including conference and journal versions, because automatically merging related publications would change the declared population. The central invariant is monotonicity: for fields enriched after the DBLP import, such as abstracts and BibTeX, the database preserves the existing non-null value whenever a later run returns null. This makes enrichment regression a violation of the storage layer, not a convention left to individual call sites.

The extraction stage fills the fields that DBLP does not provide. For each record missing an abstract, \tv queries a priority-ordered registry of open scholarly APIs and publisher-specific extractors. The first valid abstract is written back, and the remaining extractors for that record are skipped. Source-level circuit breakers and checkpoints keep long refreshes auditable: an unstable upstream can be suppressed for the remainder of a run, and an interrupted run resumes without reprocessing committed batches. The May snapshot retains canonical links but predates persistent per-record extractor provenance.

The BibTeX stage first retrieves the canonical DBLP BibTeX when available, then uses a local generator for records whose endpoints return no entry. This design keeps citation export usable even when upstream metadata is incomplete. In the released snapshot, the combined path yields 9,924 non-empty BibTeX entries out of 9,925 records.

SQLite is the sole persistent store. This design serves R6 directly: the database is a single file that can be distributed as a research artifact, committed to a repository, shared via a download link, or packaged in a container. There is no external database server to configure, no credentials to manage, and no migration step required beyond a thin schema-evolution helper invoked on each startup.

The monotonic merge is a storage-layer property, not an application convention. It combines write-with-replace semantics with a null-preserving coalesce over every enrichable field, so even application code that bypasses the high-level API cannot regress previously enriched data.

A compressed-snapshot bootstrap complements the database: the repository ships a 15~MB compressed dump of the full 74~MB database, which materializes on first use in under a second and preserves local modifications against later upstream updates. Reviewers therefore use the corpus immediately without running the pipeline, satisfying R6 with a single committed artifact rather than a multi-megabyte CSV.

Abstract extraction uses a registry of interchangeable source adapters. Each adapter returns either an abstract or no result; the orchestrator holds an ordered list of adapters per venue type, and invokes them until a result is obtained. The current set covers Semantic Scholar, OpenAlex, CrossRef, ACM Digital Library, IEEE Xplore, USENIX, and NDSS.

A new source becomes available with a single adapter and no changes to the orchestrator or the database, satisfying R2 by making it easy to add open-source components as they appear and to replace sources that become unavailable or rate-limited.

The command-line interface exposes the full construction pipeline and the corpus operations of search, export, and statistics, and can target an alternate data directory for continuous integration (CI) and scripted workflows. It is the primary interface for automation and reproducibility: read-only queries and exports are deterministic for a given snapshot, whereas network-backed construction reflects upstream changes and produces a new versioned snapshot rather than silently redefining an existing one.

The Streamlit interface (\texttt{topvenues web}) provides an interactive search with filters for venue, year range, author, and keyword, along with per-venue coverage dashboards, topic distribution charts, and one-click BibTeX and CSV exports. It adds no functionality beyond the CLI, since every search and export ultimately issues the same database queries.

\section{Reproducibility Controls, Corpus Scope and Coverage}
\label{sec:implementation}

The implementation focuses on reproducibility controls at the points where corpus drift typically occurs. Enrichment is the network-intensive stage because it calls several scholarly APIs, so it runs concurrently over configurable batches and checkpoints, with each committed batch; an interrupted run resumes without reprocessing records it has already enriched. Every record is validated against the database schema before storage, and the released gate rejects empty or implausibly short abstracts. These syntactic checks do not establish semantic correctness or exclude every form of upstream boilerplate. The manual-audit protocol treats that as a separate empirical question. The review scope, comprising venues, year ranges, and extractor priorities, lives in a versioned \texttt{config.yaml}. Selecting among already supported venues is a configuration change. Adding a new venue class also requires an explicit URL strategy and name-normalization rule and may require a source-specific extractor. This keeps scope changes auditable without implying that arbitrary venues work by configuration alone.

The tests enforce the same discipline. The suite builds small SQLite databases in temporary directories and runs offline in under 1 second. Its integration tests exercise the monotonic upsert by inserting a record with an abstract, replaying a consolidation that returns null for that field, and asserting that the original abstract survives, validating R4, which requires it to hold. A compressed SQLite snapshot ships with the repository and materializes on first run, allowing a reviewer to reproduce the corpus state without credentials or network access.


The 11-venue profile is a deliberately bounded starting population, not a census of cybersecurity. It contains the four CORE~A* venues (USENIX Security, ACM~CCS, IEEE~S\&P, NDSS)~\cite{core2023}; four adjacent venues with recurring security-relevant work (ACM~ASIA~CCS, IEEE~EuroS\&P, ACM~SACMAT, HotNets); and three broad survey venues that cover security and neighboring areas (ACM Computing Surveys, IEEE Communications Surveys \& Tutorials, Foundations and Trends in Privacy and Security). Inclusion is venue-bounded rather than topic-filtered: a record appears because it is on a configured DBLP venue page, not because every paper in that venue was classified as cybersecurity. This profile keeps scope decisions inspectable while spanning both primary studies and surveys; separately versioned profiles can broaden the collection without changing the reported denominator. The 2019 cutoff was fixed before the measurements to target recent work and provide four years of prior-publication history for the 2023 readiness cohort. It is not a claim that earlier work is irrelevant, and results should not be generalized to pre-2019 literature. Two older Foundations and Trends records exposed by the selected DBLP page are retained and reported rather than silently removed.

Table~\ref{tab:coverage} reports per-venue abstract and BibTeX coverage. Abstract coverage is at or above 99.3\% across all major venues; the single outlier is Foundations and Trends in Privacy and Security (8 papers total, 7 with abstracts), which publishes long monograph-style papers that are rarely indexed by open abstract sources. BibTeX coverage is 100\% for all venues except IEEE~S\&P (99.9\%); the single missing entry remains visible by title, authorship, venue, year, Digital Object Identifier (DOI), source web address, and DBLP key.

\begin{table}[ht]
\centering
\caption{Corpus coverage by venue (May 2026 snapshot).}
\label{tab:coverage}
\begin{tabular}{lrrr}
\toprule
Venue & Papers & Abstract & BibTeX \\
\midrule
USENIX Security Symposium        & 2,054 & 99.9\% & 100.0\% \\
ACM CCS                          & 1,972 & 99.9\% & 100.0\% \\
ACM Computing Surveys            & 1,734 & 99.9\% & 100.0\% \\
IEEE S\&P                        & 1,165 & 99.9\% &  99.9\% \\
IEEE Commun.\ Surveys            &   810 & 99.3\% & 100.0\% \\
NDSS                             &   794 & 99.9\% & 100.0\% \\
ACM ASIA CCS                     &   628 & 100.0\% & 100.0\% \\
IEEE EuroS\&P                    &   339 & 100.0\% & 100.0\% \\
HotNets                          &   245 & 100.0\% & 100.0\% \\
ACM SACMAT                       &   176 & 100.0\% & 100.0\% \\
Foundations and Trends           &     8 &  87.5\% & 100.0\% \\
\midrule
\textbf{Total}                   & \textbf{9,925} & \textbf{99.86\%} & \textbf{99.99\%} \\
\bottomrule
\end{tabular}
\end{table}

The corpus also records the publication dates of the literature. Annual volume rises from 820 papers in 2019 to 1,991 in 2025, with the most recent full conference years each contributing between 1,600 and 2,000 records. The 397 records dated 2026 are survey-journal records, 244 from ACM Computing Surveys and 153 from IEEE Communications Surveys \& Tutorials. The released snapshot contains no 2026 proceedings from the configured conferences. Two pre-2019 records remain because they appear on DBLP venue pages within the configured scope.

We also ran a live known-record comparison on July~21, 2026, using a venue-stratified sample of 200 records drawn from the 9,911 non-empty abstracts (seed 20260721). Table~\ref{tab:baseline} asks whether DBLP~\cite{ley2009dblp}, Semantic Scholar (S2)~\cite{ammar2018}, OpenAlex~\cite{priem2022}, and the local snapshot align each sampled record and expose an abstract. It is conditional on inclusion in \tv. The table measures field availability and read-path behavior, not recall for papers outside our denominator or end-to-end corpus-construction cost. Semantic Scholar used one DOI batch for 143 records and 57 paced title matches; OpenAlex used 143 DOI singletons and 57 title searches. Scripts and per-record observations ship with the artifact.

\begin{table}[ht]
\centering
\small
\setlength{\tabcolsep}{4pt}
\caption{Resolving the same 200 records from the local snapshot and from live indices. Latency is per-request median/p95; HTTP counts are logical requests.}
\label{tab:baseline}
\begin{tabular}{lrrrr}
\toprule
System & Aligned & Abstract & HTTP req. & Latency (ms) \\
\midrule
\tv (local snapshot) & 200/200 & 200/200 &   0 & 0.004/0.007 \\
DBLP                 & 200/200$^{a}$ & 0/200 & --- & --- \\
Semantic Scholar     & 199/200 & 163/200 &  58 & 1,083/2,834$^{b}$ \\
OpenAlex             & 171/200 & 161/200 & 200 & 362/381$^{c}$ \\
\bottomrule
\multicolumn{5}{p{0.95\columnwidth}}{\footnotesize $^{a}$Aligned by the DBLP keys that define the corpus, not an independent live-recall estimate; DBLP publishes no abstract field, and we therefore report no DBLP latency. $^{b}$Title-search path; the 143-record DOI batch returned in 1,237~ms as a single request. $^{c}$DOI path; the 57 title searches ran at 512/895~ms.}
\end{tabular}
\end{table}

The comparison exposes the trade-off directly. DBLP supplies stable primary metadata but no abstracts. The local snapshot resolves every sampled abstract without a network call. Semantic Scholar and OpenAlex resolve 163/200 (81.5\%) and 161/200 (80.5\%), at hundreds of requests and three orders of magnitude more latency per record. Cost is better read as quota than as money: all three services are free to query, and OpenAlex meters a daily allowance in dollar-denominated units. Its DOI lookups consume none of that allowance. The 57 title searches consumed \$0.057 of the \$0.10 daily quota, over half a day's budget for 57 records. Search-based reconciliation, not lookup, is the path that does not scale. Text agreement is a diagnostic rather than ground truth: 145 of the 163 Semantic Scholar abstracts and 120 of the 161 OpenAlex abstracts reach token-set Jaccard 0.95 against the snapshot, and the remaining disagreements need authoritative-source inspection before agreement could be read as extraction accuracy.

\section{Evaluation and Scientific Results}
\label{sec:evaluation}

We evaluate \tv against the six requirements in Section~\ref{sec:problem} using measurements taken on the released artifact, without network access, on an Apple M4 Max workstation with 64~GB of memory, Darwin 25.4.0, and Python~3.14.5. The evaluation has two layers. The first establishes artifact quality: coverage, latency, exportability, monotonic updates, and tests. The second asks whether the artifact enables a result that the community could not reliably obtain from ad hoc portal searches: early detection of top-tier cybersecurity research on arXiv. Together, these layers establish both practical utility and scientific yield: the same declared corpus that accelerates review work also becomes an auditable denominator for cybersecurity measurement.

The corpus achieves 99.86\% abstract coverage and 99.99\% BibTeX coverage over all 9,925 records (Table~\ref{tab:coverage}). Coverage counts records, including the non-research front matter that DBLP indexes for these venues: 33 records are quarterly editorials, obituaries, or a chair's message rather than papers, which the artifact flags so the denominator can be restricted to the 9,892 research papers, at 99.94\% abstract coverage. The 14 records without an abstract are six such research papers, whose open-source abstracts are not exposed, and eight front-matter records that carry none by nature, six of them 2025 IEEE Communications Surveys editorials.

The 9,910 records with both abstract and BibTeX fields provide the foundation for abstract-aware search and direct citation export. The remaining 15 records are still usable bibliographic records: one has an abstract but no BibTeX entry, and 14 have BibTeX but no abstract. This accounting makes the denominator explicit and prevents hard-to-enrich papers from being hidden from the corpus.

Table~\ref{tab:latency} reports keyword search latency for five representative cybersecurity topics. Each query performs a case-insensitive substring match against both \texttt{title} and \texttt{abstract} on the full 9,925-row table, without a full-text index, and materializes the complete result set. Latency is measured from issuing the SQLite query to receiving all result rows. All queries are completed in under 31~ms, which is imperceptible to an interactive user. Because the match uses a leading wildcard, a secondary index cannot serve it, and SQLite scans every row. Lower latency at a larger scale would require a full-text index.

\begin{table}[ht]
\centering
\caption{Keyword-search latency on the full corpus.}
\label{tab:latency}
\begin{tabular}{lrr}
\toprule
Query term & Results & Latency (ms) \\
\midrule
machine learning    & 1,079 & 20.1 \\
fuzzing             &   361 & 17.7 \\
intrusion detection &   120 & 28.1 \\
threat intelligence &    36 & 28.9 \\
ransomware          &    31 & 24.2 \\
\midrule
\multicolumn{2}{l}{Median} & 24.2 \\
\bottomrule
\end{tabular}
\end{table}

\tv supports three export formats from the CLI and web interface. Export latency was measured for the full corpus and for the post-2022 subset. BibTeX serialization for 5,922 post-2022 entries takes 2.2~ms after rows are materialized and produces a 4.4~MB \texttt{.bib} file ready for direct inclusion in a LaTeX project via \texttt{\textbackslash{}bibliography\{topvenues\}}. Full-corpus CSV export takes 213.4~ms for 9,925 rows and produces a 24~MB file that carries the bibliographic fields together with abstracts and BibTeX entries. The full-corpus JSON export takes 90.7 ms and supports programmatic downstream analysis.

The next three workflows illustrate how the same fixed corpus supports common cybersecurity research tasks without changing the denominator. In an intrusion-detection survey, a researcher issues a topic-specific query over the full corpus. The query returns 120 papers spanning 2019 to 2026 across 10 venues in 28 ms. Refining the filter to conference venues over the configured 2021 to 2026 window leaves 68 papers, each with title, authors, venue, year, and abstract; a BibTeX export of those 68 entries serializes in under 20~ms.

For CTI methodology, a researcher issues a query against the same corpus and obtains 36 papers in 29~ms, distributed across ACM CCS (12), USENIX Security (7), ACM Computing Surveys (4), NDSS (4), IEEE Communications Surveys (3), IEEE EuroS\&P (3), ACM ASIA CCS (2), and ACM SACMAT (1). This distribution contributes to survey venues immediately inspectable under one fixed denominator, whereas a multi-portal workflow would require a separate reconciliation step to assemble the same view.

For LLM security, a keyword query returns 288 papers. Their temporal distribution is one paper in 2021, ten in 2023, 73 in 2024, 152 in 2025, and 52 among the 2026-dated survey-journal records. Because the released snapshot contains no 2026 conference proceedings, the last value is not a partial estimate of the full 2026 venue profile. The distribution makes the recent concentration inspectable without manual counting across portals.

Each case study resolves the same corpus, counts, and exports from one fixed denominator in interactive time. Reaching the same result by hand means querying venue portals separately, reconciling duplicates, normalizing citations, and documenting omissions; \tv removes that source of denominator drift and leaves screening and synthesis to the researcher.

\label{sec:earlysignal}

The final case study treats the corpus as a \emph{measurement substrate}: a stable, declared population against which an external signal can be correlated reproducibly. We ask a question that a manual multi-portal workflow cannot answer with a fixed denominator: \emph{how often, and how far in advance, does top-tier cybersecurity work appear as an arXiv preprint before its formal publication?}

We take the corpus cohort of 2{,}537 papers from the four CORE~A* venues published in 2024 to 2025 and match each against a harvested snapshot of 27{,}749 arXiv preprints in the Computer Science, Cryptography and Security category (cs.CR) spanning 2022 to 2026, collected through the public arXiv API~\cite{arxiv2024api}. A preprint matches a paper when at least one author key coincides, using last name plus first initial after accent and DBLP suffix normalization, and when the normalized-token Jaccard similarity between titles is at least 0.55. Here, Jaccard similarity is the fraction of shared normalized title tokens over the union of title tokens. Because the rule favors specificity, it may miss reworded titles and changed author names. False-positive title-author coincidences nevertheless preclude treating the observed rate as a formal lower bound. The preprint-to-publication lag is anchored to the first day of each venue's conference month (USENIX~Security August, ACM~CCS October, IEEE~S\&P May, NDSS February) rather than January~1, which would otherwise create spurious negative lags for papers preprinted earlier in the publication year.

Table~\ref{tab:earlysignal} reports the outcome: 742 of 2{,}537 papers (29.2\%) have a matching preprint, with a stable rate across the four venues (28.4\% to 31.1\%). The preprint precedes the conference by a median of 154 days (roughly 5 months), with a 25th-to-90th percentile range of 47 to 442 days. The few negative lags correspond to camera-ready versions posted shortly after acceptance, not to measurement error.

\begin{table}[ht]
\centering
\caption{arXiv early-signal rate for 2024 to 2025 CORE~A* papers.}
\label{tab:earlysignal}
\begin{tabular}{lrrr}
\toprule
Venue & Papers & Preprinted & Rate \\
\midrule
USENIX Security & 857 & 243 & 28.4\% \\
ACM CCS         & 813 & 237 & 29.2\% \\
IEEE S\&P       & 516 & 153 & 29.7\% \\
NDSS            & 351 & 109 & 31.1\% \\
\midrule
\textbf{Total}  & \textbf{2{,}537} & \textbf{742} & \textbf{29.2\%} \\
\bottomrule
\end{tabular}
\end{table}

A 29.2\% coverage rate indicates that top-tier work appears early, but not \emph{which} of the ${\sim}7{,}000$ annual cs.CR preprints are worth watching. We therefore test a triage filter: does authorship by a researcher with a \emph{prior} publication in one of the four CORE~A* security venues predict that a preprint will later appear in that same four-venue scope? The test separates the predictor and the outcome. The predictor (prior-top-4 authorship) is computed strictly from publications \emph{before} the preprint year, and the outcome (``became a top-4 paper'') is decided purely by title similarity against the published corpus with no author overlap required, so the predictor and outcome are measured independently.

Table~\ref{tab:readiness} reports the result together with baselines and operating points. Requiring any prior top-4 coauthor flags 1,872 of the 5,185 preprints submitted in 2023, captures 298 of the 330 that later appear in the four-venue scope, and leaves 32 positives in the 3,313 excluded preprints. The flagged and excluded risks are therefore $298/1{,}872=15.9\%$ and $32/3{,}313=0.97\%$: their ratio is a 16.5$\times$ \emph{relative risk} (RR), not lift. Conventional lift divides precision by population prevalence, giving $15.9\%/(330/5{,}185)=2.5\times$. The filter is tunable by author position: first-author history raises precision to 26.4\% at 42\% recall, while senior-author history lies between. The association is not explained by publication volume or community membership alone: prolific-author, any-security-author, and seeded random-author controls show different operating points. It also replicates on the 2022 cohort and is stable across title-match thresholds from 0.5 to 0.7.

\begin{table}[ht]
\centering
\small
\setlength{\tabcolsep}{4pt}
\caption{Scientific-readiness triage for 5,185 2023 cs.CR preprints (330 positives). RR compares flagged with excluded risk; lift uses population prevalence.}
\label{tab:readiness}
\begin{tabular}{lrrrrr}
\toprule
Filter & Hit/flagged & Prec. & Recall & RR & Lift \\
\midrule
Prior top-4 author (any)    & 298/1,872 & 15.9\% & 90.3\% & 16.5$\times$ & 2.5$\times$ \\
\quad first author only     & 138/523   & 26.4\% & 41.8\% &  6.4$\times$ & 4.1$\times$ \\
\quad senior (last) author  & 237/1,176 & 20.2\% & 71.8\% &  8.7$\times$ & 3.2$\times$ \\
Any security-venue author   & 308/2,410 & 12.8\% & 93.3\% & 16.1$\times$ & 2.0$\times$ \\
Prolific ($\geq$3 papers)   & 249/1,338 & 18.6\% & 75.5\% &  8.8$\times$ & 2.9$\times$ \\
Random security authors     & 256/1,758 & 14.6\% & 77.6\% &  6.7$\times$ & 2.3$\times$ \\
\bottomrule
\end{tabular}
\end{table}

Both results are reproducible \emph{because} the corpus is a declared population. The denominators are fixed and auditable, the measurement procedures ship with the artifact and are exercised by dedicated tests, and the results regenerate from a one-time arXiv harvest followed by fully offline analysis against the cached snapshot. Together they turn literature monitoring into a repeatable measurement, and give surveillance work a practical triage strategy for a preprint volume that defeats manual inspection.

The test suite contains 252 tests across 15 modules, all passing on the released artifact in under 1 second without network access, and continuous integration re-runs them on every commit across Python 3.11 and 3.12. Testing focuses on the components most likely to corrupt the corpus silently: record validation, venue normalization and deduplication, abstract extraction, and the database layer that enforces the \texttt{COALESCE} monotonicity invariant. The full per-module breakdown ships with the artifact.

\section{Discussion}
\label{sec:discussion}

The submission snapshot is bound to the SHA-256 digest \texttt{0f4dbaa9\ldots ef64cd} and an offline audit of the exact bytes makes the provenance boundary explicit. Of the 9,911 non-empty abstracts, 8,525 match retained enrichment-log or API-cache evidence exactly after whitespace normalization. Another 1,386 have no retained source evidence, and 14 records have no abstract. An exact match is evidence consistent with a retained source, not proof of which extractor wrote the field, because the May snapshot predates persistent per-record extractor provenance. Abstract coverage in this paper therefore means validated non-empty text, not publisher-canonical provenance. The released pipeline has no PDF-text extraction stage, and the historical number of PDF-derived abstracts cannot be reconstructed from the snapshot; we report it as unknown rather than infer zero. The audit manifest preserves this uncertainty instead of assigning sources retrospectively.

The corpus separates two kinds of stability that are often conflated in literature review tooling. The first is denominator stability: the set of records in a released snapshot must remain fixed so that a review or measurement can be cited, rerun, and compared later. The second is metadata quality: abstracts, source web addresses, and BibTeX entries can improve as upstream sources change or as users report defects. \tv handles this tension by treating the snapshot as immutable evidence and the construction pipeline as a repeatable process. A published snapshot is never silently rewritten; a later release can replace a lower-quality field with one with a better provenance path, but the change is visible in the database digest, the lineage metadata, and the release notes. This policy preserves the value of monotonic enrichment for bulk refreshes while still allowing explicit corrections when a field is wrong, truncated, or associated with the wrong paper.

This separation is the paper's core design insight. A literature review involves at least three decisions, often collapsed into a single spreadsheet: which population is eligible, which records are relevant to the research question, and which evidence the authors synthesize from those records. \tv does not automate the last two decisions. Instead, it makes the first one a reproducible object with a name, a scope, a digest, and recorded source evidence. That boundary is valuable because it localizes disagreement. If two reviewers disagree about whether a paper belongs in an intrusion-detection survey, they can now determine whether their disagreement concerns the declared population or the paper's topical relevance within that population. Without that separation, both disputes are mixed with portal drift, missing abstracts, duplicate records, and citation-cleanup choices.

The unit of analysis is a DBLP bibliographic record, not an inferred scholarly-work cluster. The released snapshot contains 9,925 distinct DBLP identifiers, no duplicate DBLP key or exact DOI, and five pairs with identical normalized title, authors, year, and pages but distinct DBLP keys and DOIs. We retain all ten records: silently collapsing the five candidate pairs would alter the accepted denominator, while title similarity alone cannot establish identity. Conference papers and journal extensions likewise remain distinct publications; a future schema can expose an explicit version relation without deleting either record. This identity boundary complements the paper's core separation between population construction, relevance judgment, and evidence synthesis: \tv fixes the first and leaves the latter two to reviewers.

The same model supports responsible redistribution. The artifact stores bibliographic metadata, recovered abstracts, stable identifiers, and links to publisher or venue pages; it does not redistribute full papers or claim ownership over their text. The source-evidence manifest distinguishes records supported by retained logs or caches from records whose historical extractor is unresolved. This keeps the data-management boundary clear: the snapshot is a reproducible index and measurement substrate, not a substitute for the publisher record, and any correction or takedown must produce an explicit versioned change.

The submission snapshot and the evolving tool serve different scientific roles. The May 2026 snapshot serves as the denominator for all numbers in this paper; subsequent data refreshes may add newly indexed papers, improve abstracts, or broaden the venue profile, but they must be designated as new releases before they can be used to support a new measurement. This distinction prevents freshness from becoming an uncontrolled variable. A researcher who wants the latest corpus can update the artifact, while a reader who wants to verify this paper can replay the released snapshot and obtain the same counts. Treating updates as versioned releases also turns future changes into auditable deltas: added papers, corrected abstracts, and altered scope profiles can be inspected without changing the evidence base behind a published claim. This release discipline is what lets \tv be both a usable tool and a reproducible measurement substrate.

SQLite fits the current single-user, read-heavy workload: 9,925 records remain small enough to mirror, search, and export interactively. The scale boundary is explicit. A corpus with hundreds of thousands of records can add SQLite Full-Text Search (FTS5); a concurrent-write service can replace the storage adapter while preserving stable DBLP identity, explicit version relations, and monotonic updates. The contribution is the declared corpus and construction contract, not SQLite itself.

Every reported count is consequently a statement over an inspectable population. The 36 CTI results, 742 arXiv matches, and readiness operating points refer to records that a reader can enumerate, including exclusions and missing fields. Fixing the published population before comparing an external preprint signal prevents source drift from masquerading as a trend. The artifact supports inspectable scope, monotonic bulk enrichment, quantified retained source evidence, and executable measurements; it does not support universal venue coverage, publisher-canonical text, complete historical attribution, or a causal effect of author history. \tv is therefore a reproducible substrate for review and monitoring methods, not a ranking or autonomous screening model.

\section{Threats to Validity}
\label{sec:threats}

The measurement environment bounds internal validity. We measured latency in Section~\ref{sec:evaluation} on a single machine; actual latency will vary with hardware. Each reported latency is the median of five runs. The SQLite storage engine's page cache may warm during repeated queries, slightly reducing apparent latency for later queries in a session.

The declared scope bounds external validity. We report measured artifact operations rather than human-time comparisons: end-to-end review time varies with the researcher, portal, topic, and amount of citation cleanup, so the manual-workflow contrast identifies sources of variance rather than a timed benchmark.

Construct validity depends on how metadata completeness and preprint matching are operationalized. Abstract coverage is the fraction of records with a non-empty \texttt{abstract} field, not a claim that every string is complete. The released validation rejects whitespace-only and short values. To measure the remaining gap, an author labeled a venue-stratified sample of 200 records against the publisher sources. Of these, 168 carry a complete and uncontaminated abstract, an intrinsic validity of 84.0\% with an exact 95\% interval of 78.2\% to 88.8\%. The failures are concentrated rather than diffuse: USENIX Security reaches 34.1\% while every other venue falls between 87.5\% and 100\%. The cause is an extractor that captured only the opening paragraph of multi-paragraph abstracts, since corrected and verified against the manual labels on every affected record. The released snapshot keeps the original text so that its digest remains citable.

The 29.2\% early-signal rate is an observed match rate, not a formal lower bound. Omission pressure comes from reworded titles, changed names, and the cs. CR-only harvest of arXiv. False-positive title-author coincidences can instead inflate the rate. As diagnostics, 695/742 (93.7\%) best matches have exactly normalized titles, and 732/742 (98.7\%) have Jaccard similarity at least 0.7. These concentrations support but do not replace manual adjudication. The lead-time anchor is the first day of each venue's conference month, an approximation accurate to within a few weeks of the actual presentation date; it is used only for the lag distribution, not for the match rate, so the headline percentage is unaffected.

The 16.5$\times$ value is a \emph{relative risk} between flagged and excluded preprints. The conventional lift over population prevalence is 2.5$\times$. Both quantify a correlational triage rule, not a causal effect of track record on acceptance. The conversion outcome is detected by title similarity against the corpus, which can miss a preprint whose title was substantially rewritten before publication. Although the protocol permits outcome years through 2026, the released snapshot contains no 2026 proceedings from the four outcome venues, so observed follow-up ends with their 2025 proceedings. Later conversions are right-censored. The operating points are conditional on this effective observation window, venue scope, and matching rule.

\section{Related Work}
\label{sec:related}

Scholarly metadata and snapshot systems provide the raw material for \tv, but not the same research object. DBLP is the closest fit for a stable computer-science bibliography: it exposes venue pages, persistent keys, and structured metadata through public interfaces~\cite{ley2009dblp,ley2002}. Semantic Scholar, OpenAlex, the Semantic Scholar Open Research Corpus (S2ORC), CrossRef, and publisher APIs provide broader metadata and, in many cases, abstracts or citation links~\cite{ammar2018,priem2022,lo2020s2orc}. \tv uses these systems as upstream sources in a controlled construction process. Their design goal is broad scholarly discovery or large-scale graph construction, whereas ours is a declared cybersecurity denominator that can be cited, inspected, and recomputed. This distinction matters for a venue-bounded review: a fresh query to a broad index may improve over time, but the improvement changes the population unless the slice, omissions, and provenance are captured in a versioned artifact. Broad scholarly graphs serve as upstream evidence sources for \tv. They help recover metadata, but they do not by themselves define which security venues, years, paper classes, and enrichment decisions constitute the population of a review.

Review tools and reporting standards address a different layer of the workflow. Reference managers such as Zotero~\cite{zotero2006} organize citations after collection, and systems such as Rayyan~\cite{ouzzani2016rayyan} and ASReview~\cite{van2021asreview} help reviewers screen or prioritize a candidate set. Reporting standards describe how to make review protocols, screening decisions, and bibliometric analyses more transparent \cite{kitchenham2007slr,page2021prisma,rethlefsen2021prismas}. Automation surveys and bibliometric guidance explain which review steps can be assisted by tools and which choices remain methodological \cite{van2021automation,linnenluecke2020conducting}. \tv sits before those tools. It makes the candidate population explicit enough that downstream relevance judgments have a stable input. This position is also why the artifact exports BibTeX, CSV, and JSON instead of replacing existing review environments.

\tv therefore occupies a different layer than the screening and reporting tools used after a candidate set is created. A review protocol can say which databases and queries were used, and a screening tool can record inclusion decisions. Still, neither artifact necessarily preserves the exact population over which those decisions were made. This gap matters when the same research question is revisited months later, because changes in source coverage can appear to be changes in the field. The contribution of \tv is to make the candidate population itself an executable object before relevance judgments begin. That design makes later screening, synthesis, and disagreement more interpretable: the reviewer can separate a dispute about scope from a dispute about relevance.

The distinction is especially important for cybersecurity because the field's evidence base is heterogeneous. A review may combine top-conference papers, survey journals, arXiv preprints, vendor reports, public feeds, vulnerability databases, and code artifacts. Tools that optimize screening effort are valuable once the candidate set exists. Still, they do not decide whether a late DBLP update, a missing publisher abstract, a renamed venue page, or a cross-listed preprint changed the denominator. \tv contributes a narrower object: a citable population for scholarly literature, with enough provenance to explain what was included, what remained incomplete, and which measurements were computed over that population.

Security reproducibility and data stewardship motivate the artifact boundary. Recent security work shows that the availability of artifacts and data remains difficult to evaluate in practice. Studies of machine-learning security papers measure gaps in code and dataset availability~\cite{olszewski2023reproducibility}, work on security data sharing shows that reusable datasets remain hard to locate and interpret~\cite{crowder2025datainfinity}. A USENIX Security Systematization of Knowledge (SoK) argues that replicability requires explicit reasoning about the problem, domain, method, data, and analysis~\cite{olszewski2025replicability}. \tv applies this artifact discipline to literature denominators. Its data-management boundary is deliberately narrow: it stores bibliographic metadata, abstracts, provenance, and reproducible measurements, while pointing back to canonical publisher or venue pages for the papers themselves. The Findable, Accessible, Interoperable, and Reusable (FAIR) principles provide the broader open-science frame for this choice~\cite{wilkinson2016fair}.

\section{Conclusion}
\label{sec:conclusion}

\tv turns corpus construction from a transient search procedure into a reproducible system boundary for cybersecurity literature reviews. The released artifact fixes the scope, preserves enriched fields through a monotonic SQLite snapshot, quantifies its source-evidence boundary, and makes the same denominator available to search, export, testing, and measurement. The May 2026 snapshot provides near-complete abstract and BibTeX field coverage across 9,925 DBLP records and reproduces the arXiv early-signal and correlational readiness-filter measurements without using live portals. This makes the population auditable before screening.

\section*{Acknowledgments}

This work is funded by FAPESP \#2020/09850-0, \#2024/21006-1, \#2022/00741-0; CNPq 312294/2025-5, 409743/2025-9; and CAPES.

\bibliographystyle{sbc}
\bibliography{references}

\end{document}